\documentclass[
    ,final            
  ]
  {aipproc}

\layoutstyle{8x11double}

\begin{document}

\title{Star Formation in Massive Low Surface Brightness Galaxies}
\classification{}
\keywords      {}

\author{K. O'Neil}{
  address={NRAO, PO Box 2, Green Bank, WV 24944}
}

%

\begin{abstract}
Massive low surface brightness galaxies have disk central surface brightnesses at least one magnitude fainter than the night sky, but total magnitudes and masses that show they are among the largest galaxies known. Like all low surface brightness (LSB) galaxies, massive LSB galaxies are often in the midst of star formation yet their stellar light has remained diffuse, raising the question of how star formation is proceeding within these systems. HI observations have played a crucial role in studying 
LSB galaxies as they are typically extremely gas rich. In the past few years we have more than quadrupled the total number of massive LSB galaxies, primarily through HI surveys. To clarify their
structural parameters and stellar and gas content,  we have undertaken a multi-wavelength study of
these enigmatic systems. The results of this study, which includes HI, CO, optical, near UV, and far UV images of the galaxies, will provide the most in depth study done to date of how, when, and where star formation proceeds within this unique subset of the galaxy population.
\end{abstract}

\maketitle


\section{Introduction}

Although Low Surface Brightness (LSB) galaxies, those objects with a central surface
brightness at least one magnitude fainter than the night sky, are now well established as
a real class of galaxies with properties distinct from the High Surface Brightness (HSB)
objects that define the Hubble sequence, considerable uncertainty still exists as to both the range of their properties and their number density in the z$\le$0.1 Universe. As LSB galaxies encompass many of the `extremes' in galaxy properties, gaining a firm understanding of LSB galaxy properties and number counts is vital for testing galaxy 
formation and evolution theories, as well as for determining the relative amounts of 
baryons that are contained in galaxy potentials compared to those that may comprise the
Intergalactic Medium, an issue of increasing importance in this era of precision cosmology.

The `traditional' (but erroneous) perception of LSB galaxies is that they are like young
dwarf galaxies: low mass, fairly blue systems, with relatively high M$_{HI}$/L$_B$ values and low
metallicities. In practice, however, LSB disk galaxies are now known to have a remarkable
diversity in properties, including very red objects, galaxies with near-solar luminosity, and
high M$_{HI}$ ($\ge 10^{10}$ M$_\odot$) systems.  LSB galaxies also include Malin 1 -- the
largest disk galaxy known to date. While none of these results contradict the idea that the
average LSB galaxy is less evolved than the average HSB galaxy, they do show that we
have not yet come close to fully sampling the LSB galaxy parameter space. In addition, it
should be emphasized that there may still be large numbers of LSB galaxies with properties
beyond our present detection limits.

Here we discuss an ongoing project to determine the properties of LSB galaxies at the massive end of the
spectrum.  The galaxies described herein all have M$_{B} \le -18$ and M$_{HI} \ge 10^9 M_\odot$.
Thus while the galaxies are not all as impressive as Malin 1, none of the galaxies could be considered
dwarf systems and quite a few may indeed be some of the largest and/or most massive galaxies known.

\section{How Many Are There?}
Since the discovery of Malin 1 \citep{bothun87}  a number of papers have been published describing the discovery of massive LSB systems \citep[e.g.][]{bothun90, sprayberry93, walsh97, davies88, sprayberry95}.  Yet until a few years ago the total number of massive LSB galaxies known was only $\sim$18.  A recent HI survey of known galaxies without known HI properties by \citet{oneil04} doubled the total number of massive LSB galaxies known ($\sim$35). 

Following up from the \citet{oneil04} survey, \citet{oneil08} have undertaken to observe the 21-cm lines of all galaxies listed in the HyperLEDA catalog with a high probability of being massive LSB systems \citep{oneil08}.  In all  257 galaxies were observed using the Arecibo, Nan\c{c}ay, and Green Bank radio telescopes.  Of these 144 galaxies had unambiguous detections, and 20 fall into the category of very massive LSB galaxies (M$_{HI} \ge 10^10$ M$_\odot$ and/or W$_{20,uncorrected} \ge $ 400 km s$^{-1}$).

\section{Optical Morphology}
\begin{figure}
\includegraphics[height=.2\textheight]{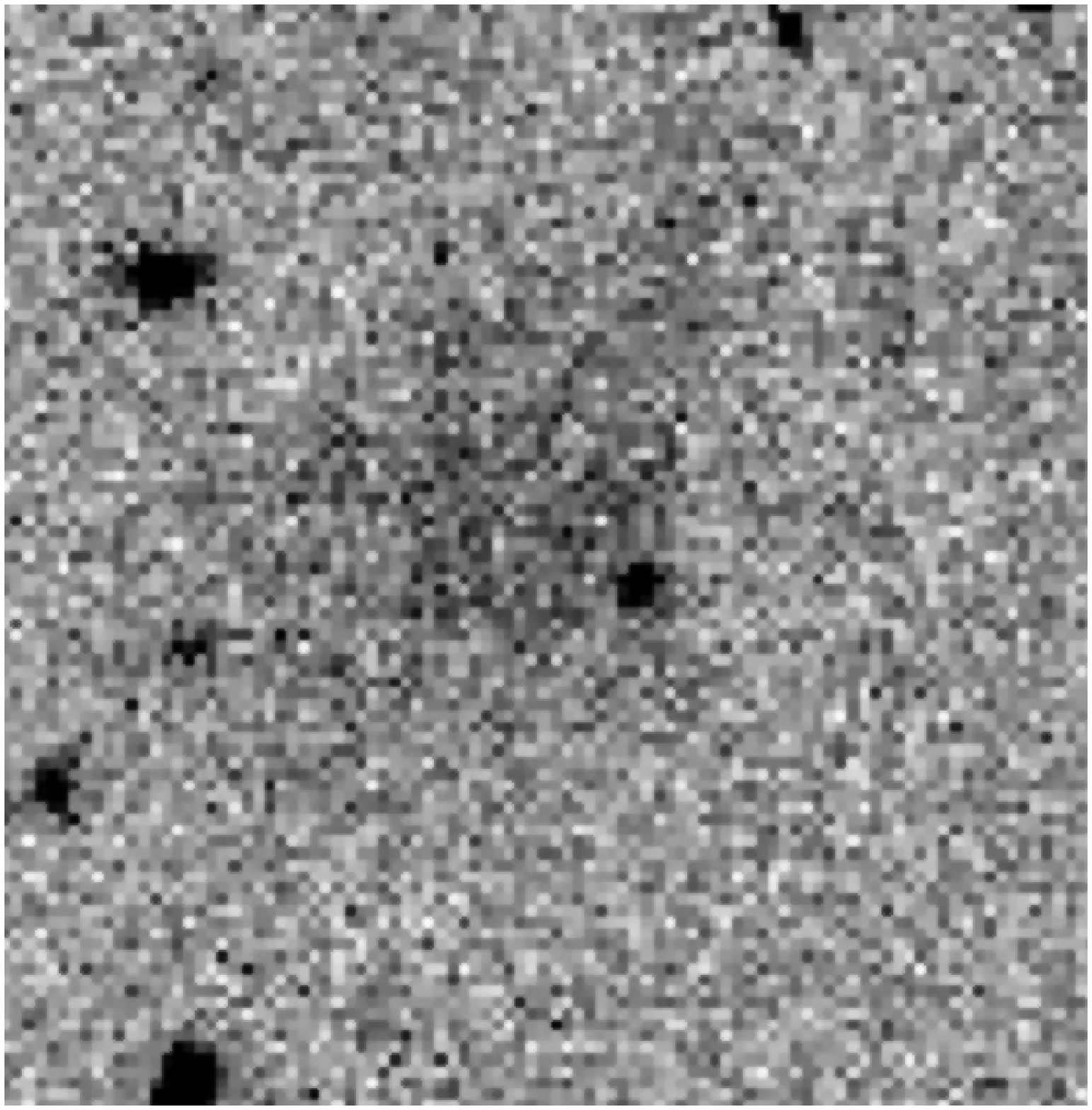}\\
\includegraphics[height=.2\textheight]{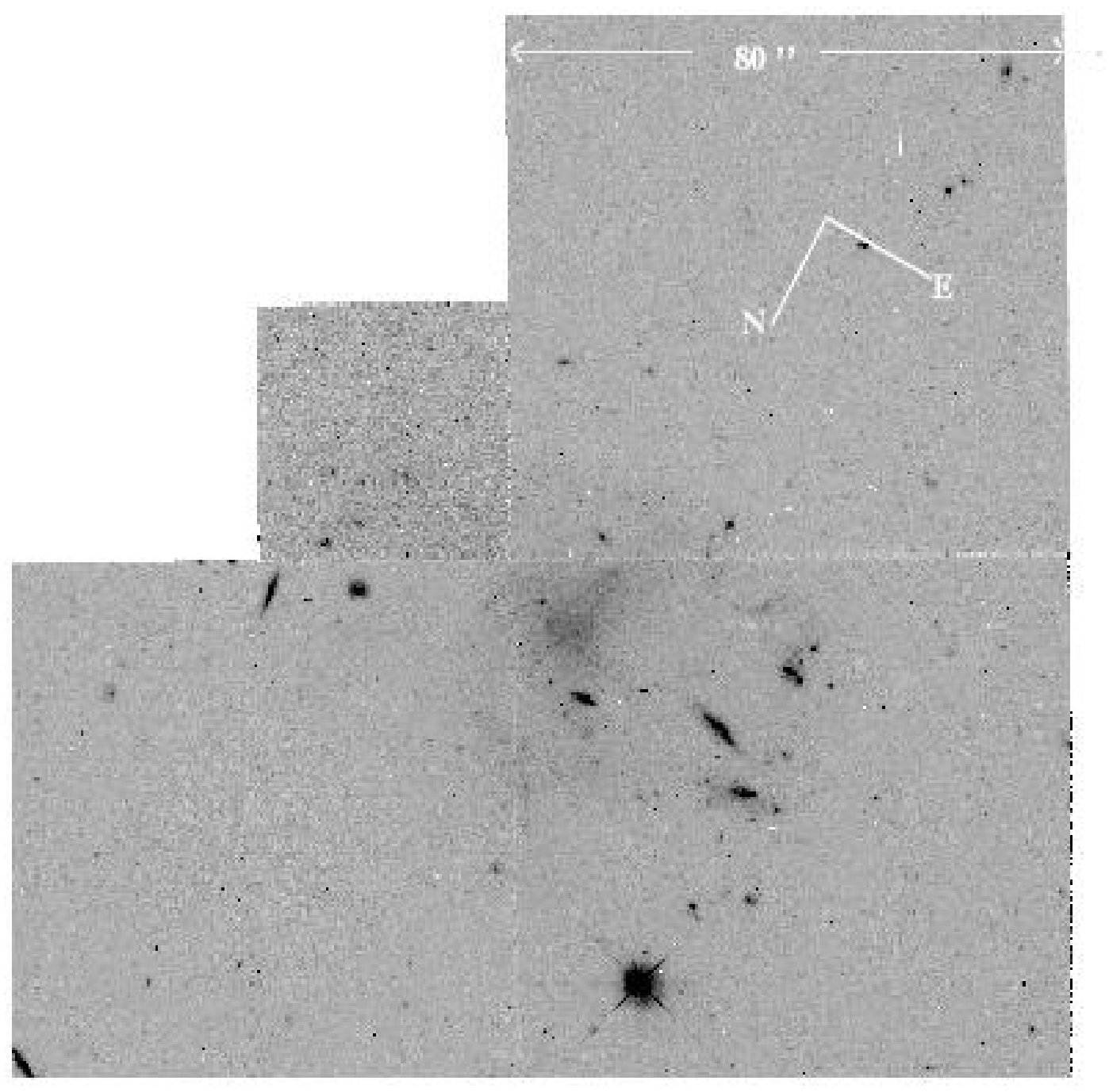}\\
\includegraphics[height=.2\textheight]{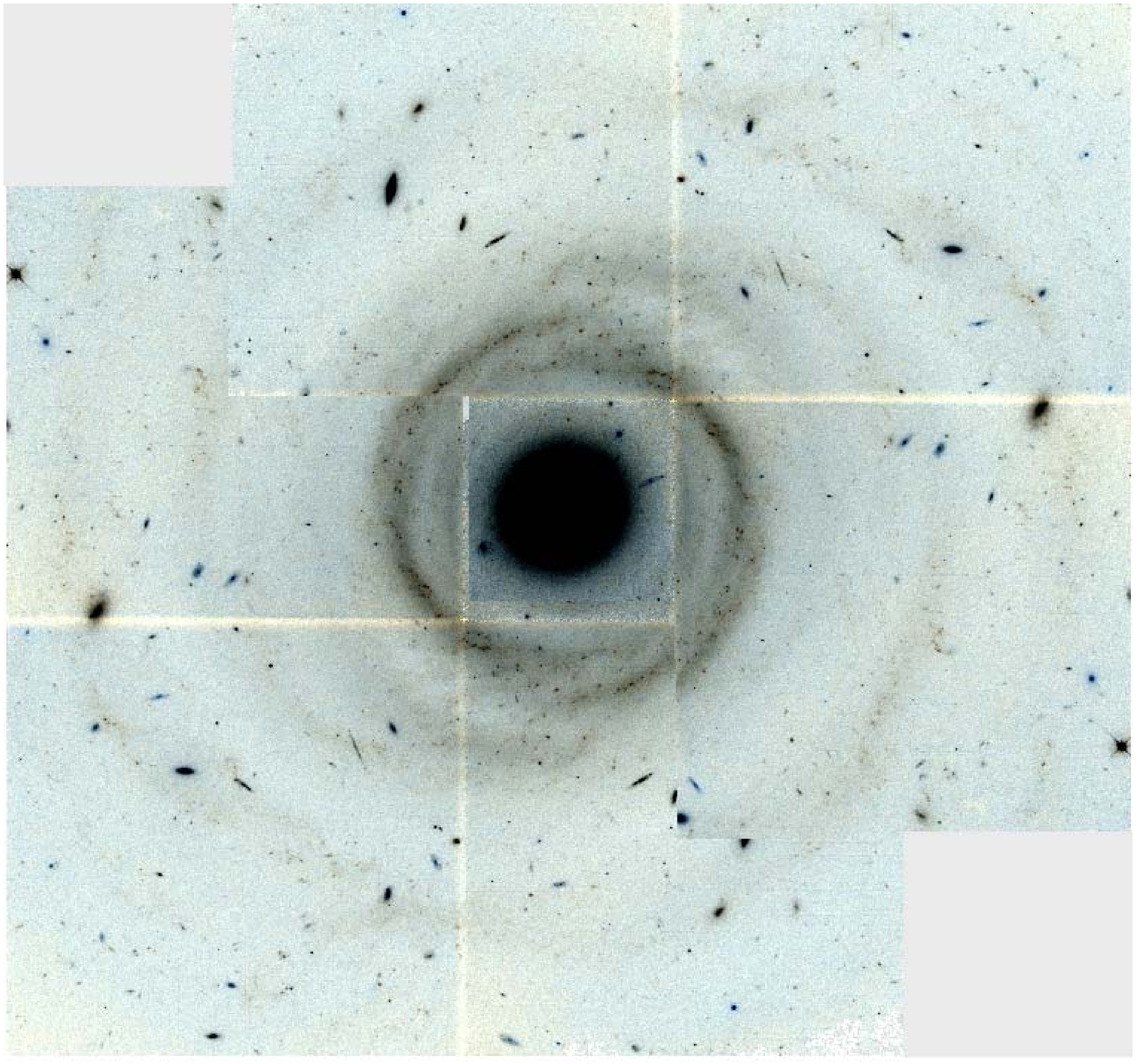}
\caption{Optical images of three LSB galaxies. Listed from left to right, and also in order
of increasing HI mass, the galaxies are [OBC97] P02-4, UGC 12695, and UGC 06614. \label{fig:morph}}
\end{figure}

The general appearance of massive LSB galaxies shows a prominent central bulge surrounded by distinct yet diffuse spiral arms (Figure~\ref{fig:morph}).  Overall the galaxies are typically less amorphous than their less massive counterparts, presumably due to the higher gravitational potential at their cores, but are still less well defined than their high surface brightness counterparts.

\section{Atomic Gas -- HI}

\begin{figure}
\includegraphics[height=.3\textheight]{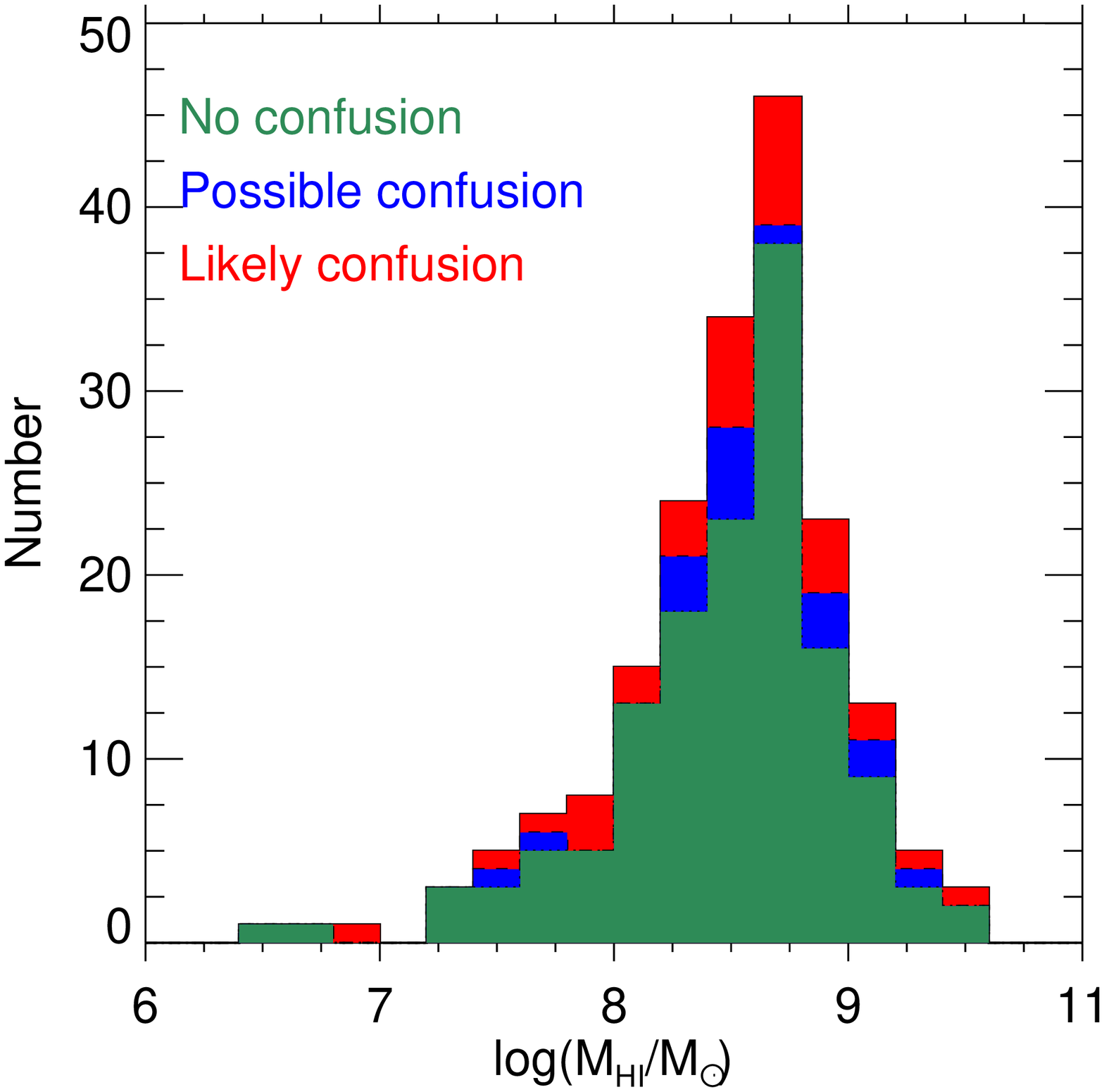}
\includegraphics[height=.3\textheight]{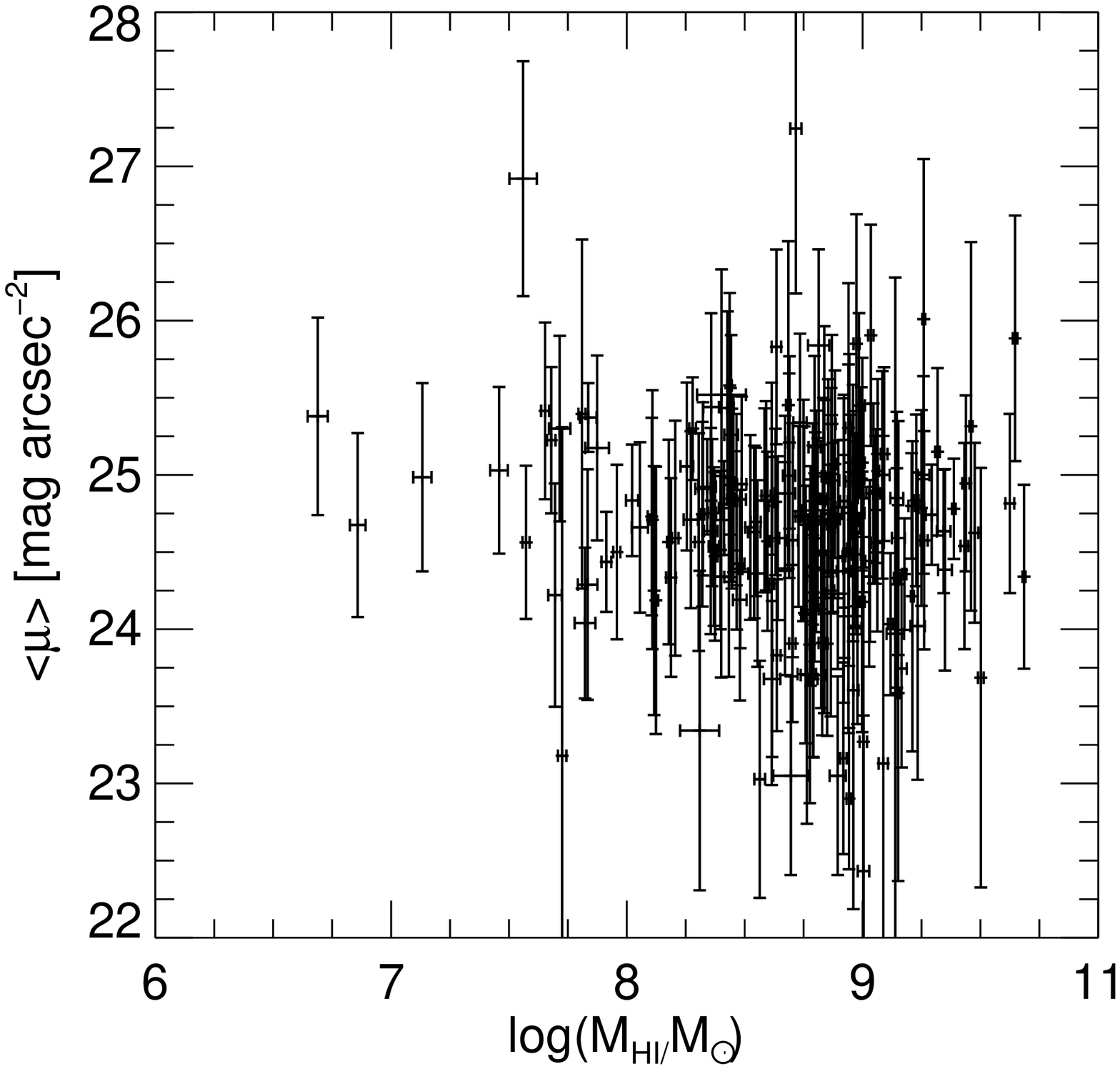}
\caption{Distribution of HI mass for the \citet{oneil08} survey.  Note that observations 
of the galaxies marked as 'confused' in the left plot likely picked up more than one galaxy 
in the telescope beam, rendering these detection highly suspect. On the right the HI mass
is plotted against average surface brightness for the galaxies.  \label{fig:HI}}
\end{figure}

As mentioned earlier, the HI mass of LSB galaxies covers the full spectrum from $<10^8$ M$_\odot$
through $>10^{10}$ M$_\odot$.  As an example, Figure~\ref{fig:HI} shows the distribution of
HI mass for the sample of galaxies observed by \citet{oneil08}.  Note that in this case the
galaxies were chosen to lie away from the dwarf galaxy realm and so the fall-off in the distribution
at the low mass end of the figures is purely a selection effect.  The same Figure also 
shows the mass distribution of sources with surface brightness, showing no trends toward higher (or lower) 
surface brightness galaxies having higher HI masses.

\section{Star Formation -- H$\alpha$ and UV Light}

\begin{figure}
\includegraphics[height=.3\textheight]{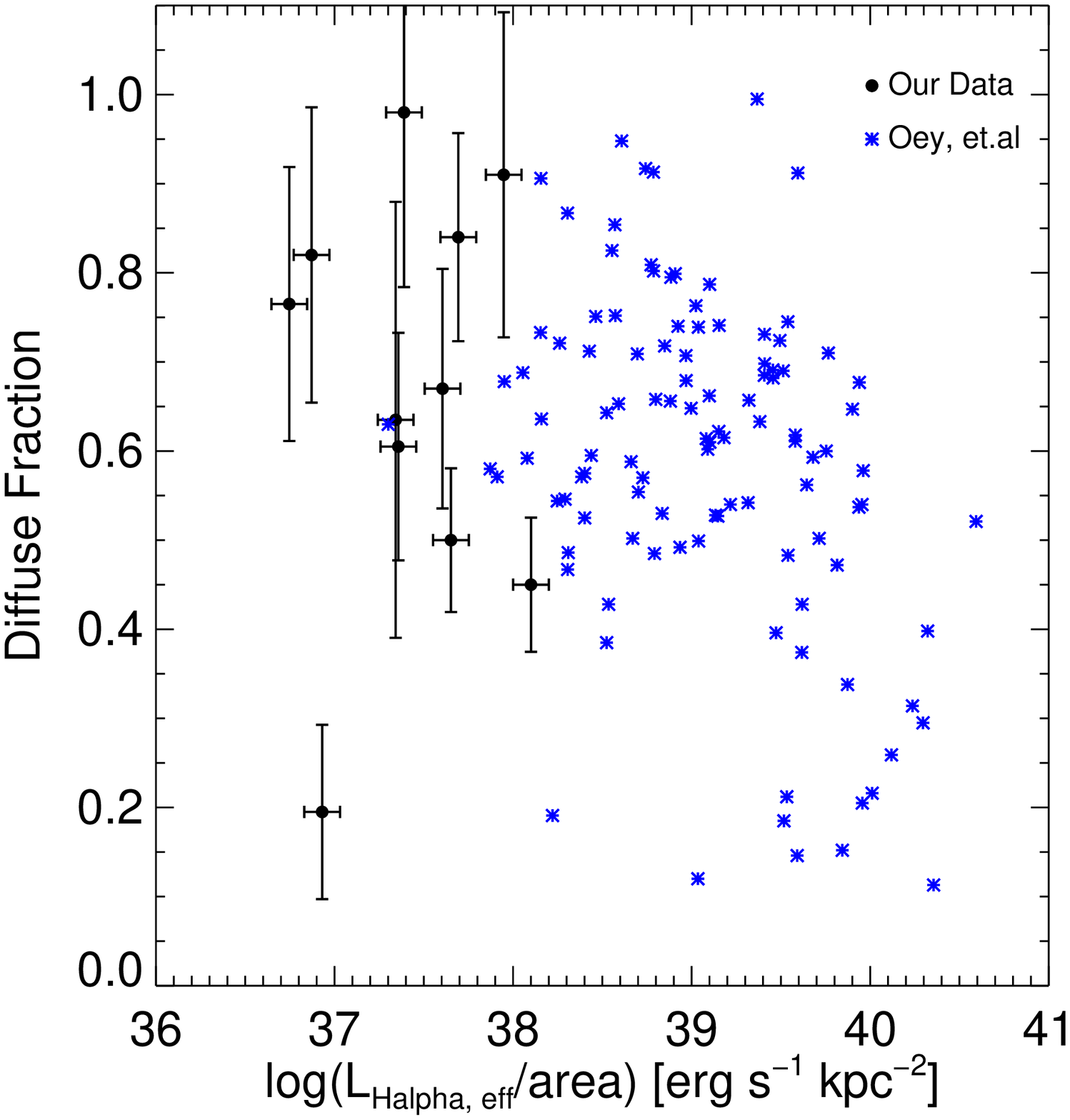}
\includegraphics[height=.3\textheight]{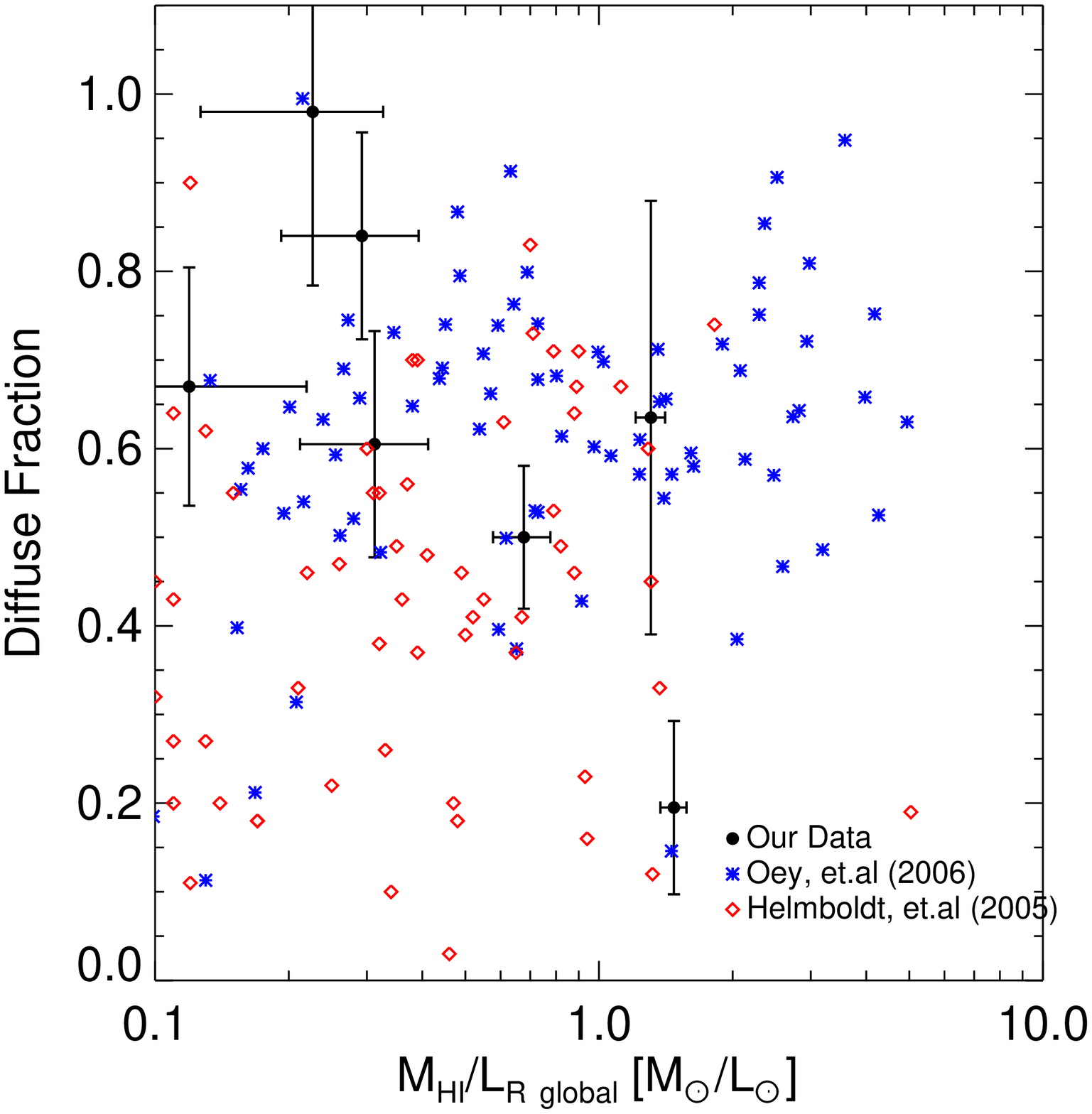}
\caption{Luminosity surface brightness (luminosity/area) (left) and gas mass-to-luminosity ratio (right) 
plotted against the diffuse H$\alpha$ sample for the sample of \citet{oneil07} ("Our Data") and that 
of \citet{oey07} and \citet{helmboldt04}.  \label{fig:halpha}}
\end{figure}

A number of studies have been done on looking at the star formation rates in LSB galaxies
\citep[e.g.][]{oneil07, helmboldt05, gerritsen96, mcgaugh94}.  Overall, the massive LSB galaxies 
appear to be forming stars at a rate similar to that of their higher surface brightness counterparts.
What is intriguing, though, is the finding of \cite{oneil07} which showed their sample to
have higher fractions of diffuse H$\alpha$ gas 
than their high surface brightness counterparts (Figure~\ref{fig:halpha}),
showing a high fraction of the ionizing photons in massive LSB galaxies lie outside the
density-bounded HII regions.

\begin{figure}
\includegraphics[height=.3\textheight]{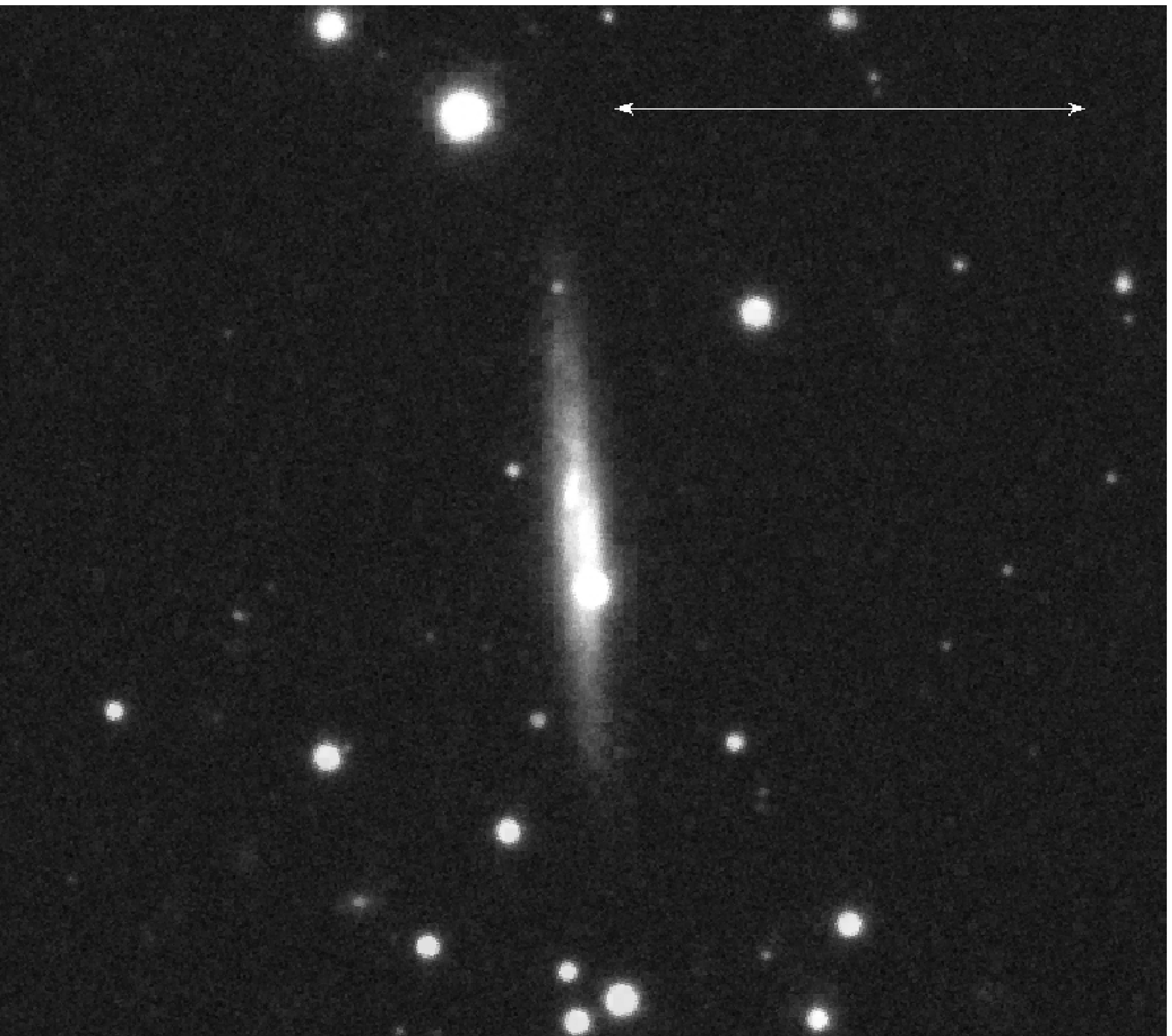}
\includegraphics[height=.3\textheight]{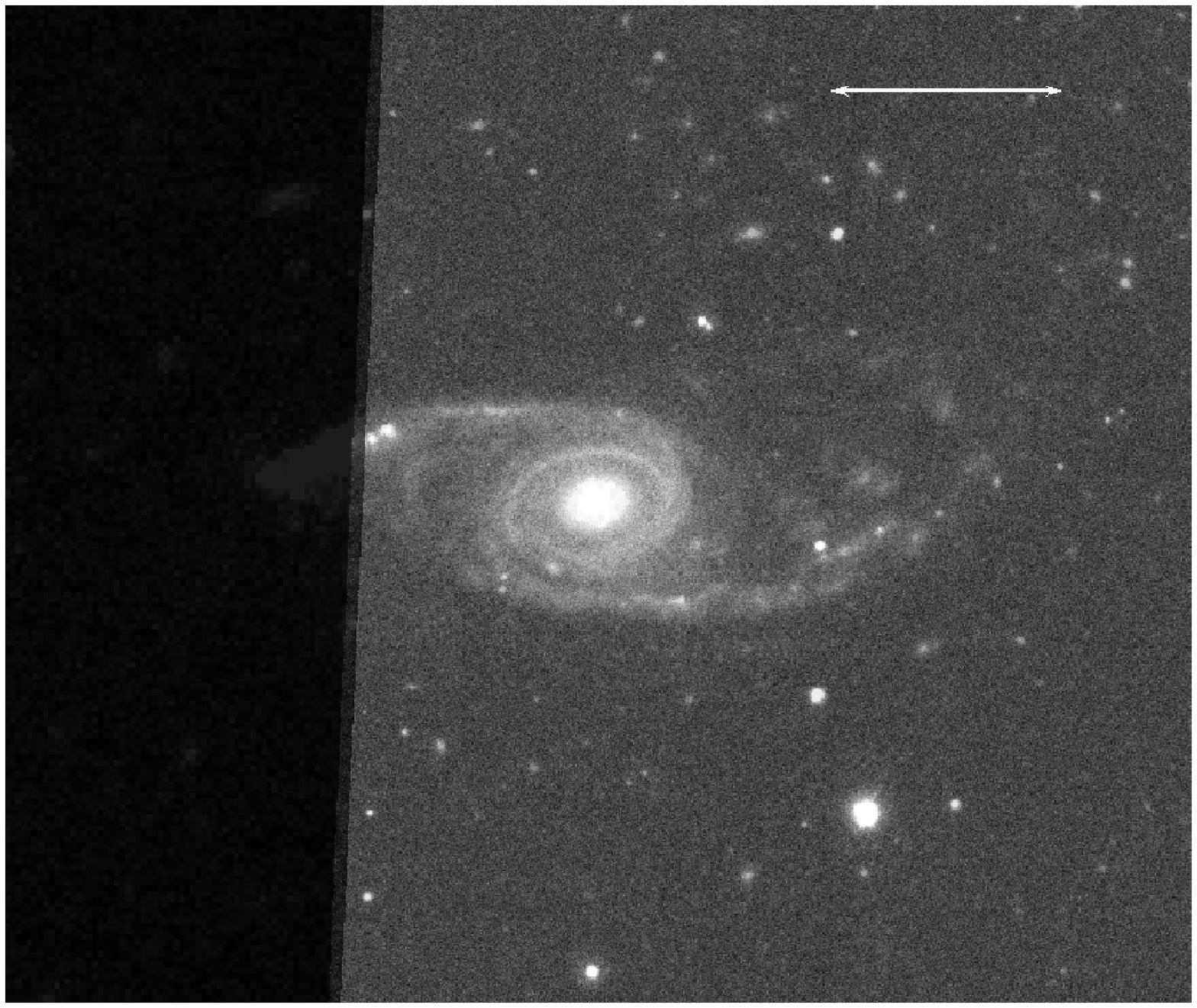}
\caption{False color images showing UGC 04144 (left) and UGC 06968 (right).  
The red and green image colors are the i and g bands from the Sloan Digital Sky Survey
\citep{stoughton02} while the blue image band is from Galex NUV images.  \label{fig:galex}}
\end{figure}

To investigate the phenomenon of star formation in massive LSB galaxies further, we have obtained
GALEX UV images of number of these systems.  Figure~\ref{fig:galex} shows a few of the galaxies
observed for this program. In agreement with the H$\alpha$ study, the Figure shows
indications of star formation which do not appear to be contained solely to the various star forming 'knots' seen in the optical images.

\section{Dust -- The Infrared}

\begin{figure}
\includegraphics[height=.2\textheight]{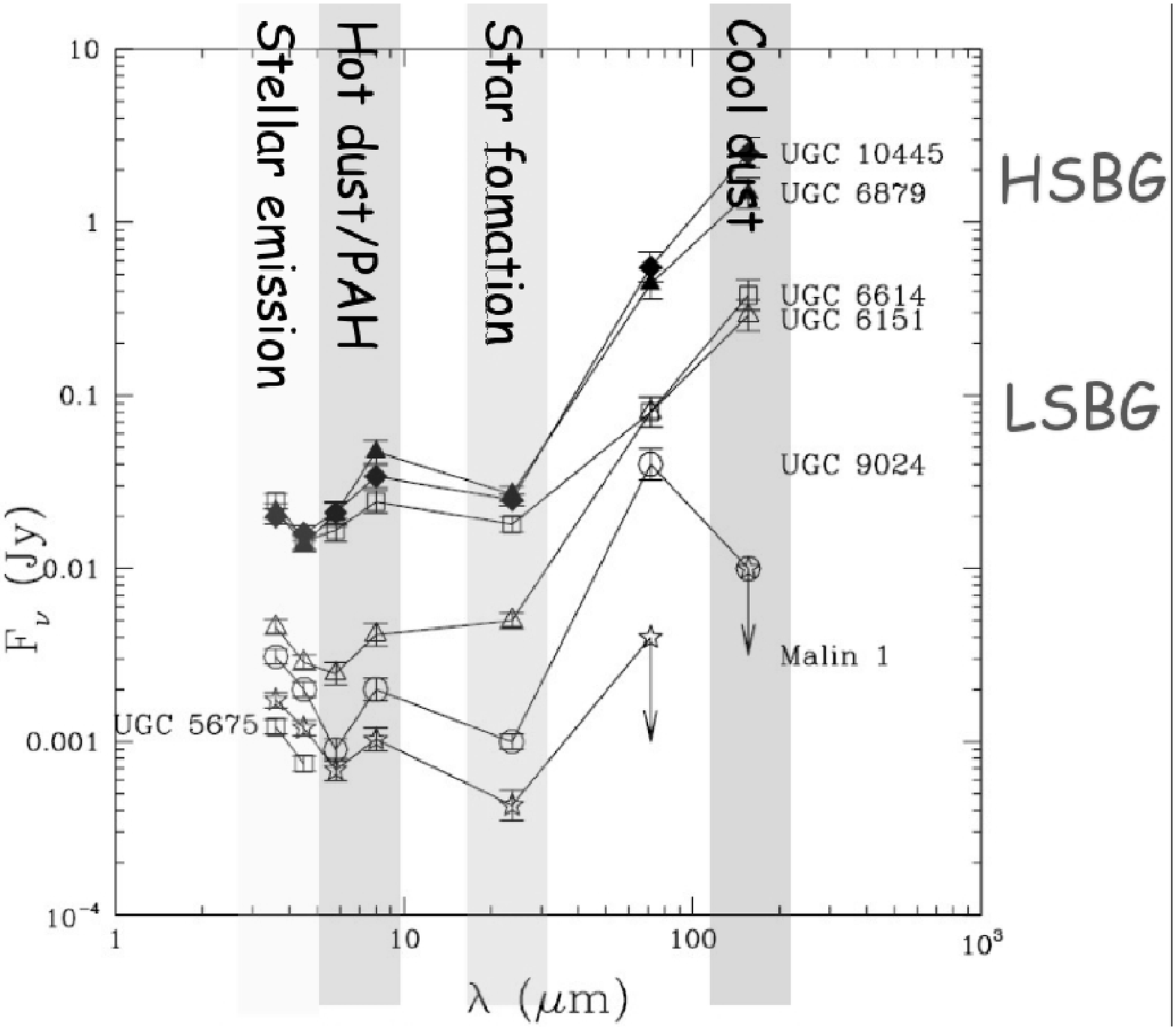}
\caption{SEDs of all the galaxies in \citet{hinz07} showing the IRAC and MIPS data points. 
The high surface brightness galaxy data are shown with filled symbols, while the LSB galaxy data 
are shown with open symbols. The arrows represent 3$\sigma$ upper limits at 70 and 160 $\mu$m.  \label{fig:dust}}
\end{figure}

Recently \citet{hinz07} obtained Spitzer observations of five low surface brightness galaxies, 
two of which are massive LSB systems (Figure~\ref{fig:dust} -- UGC 06614 and Malin 1 are the
massive LSB galaxies which were studied).  
Stellar emissions, hot dust, and aromatic molecules
were detected from all observed galaxies with uncorrupted data (the 24 $\mu$m data from one
galaxy was unusable).  At the 70$\mu$m and 160$\mu$m wavelengths, where cool dust would
be found, only two of the galaxies were detected, with the strength of the dust emission
apparently dependent on the existence of bright star forming regions.

\section{Molecular Gas -- CO}

\begin{figure}
\includegraphics[height=.3\textheight]{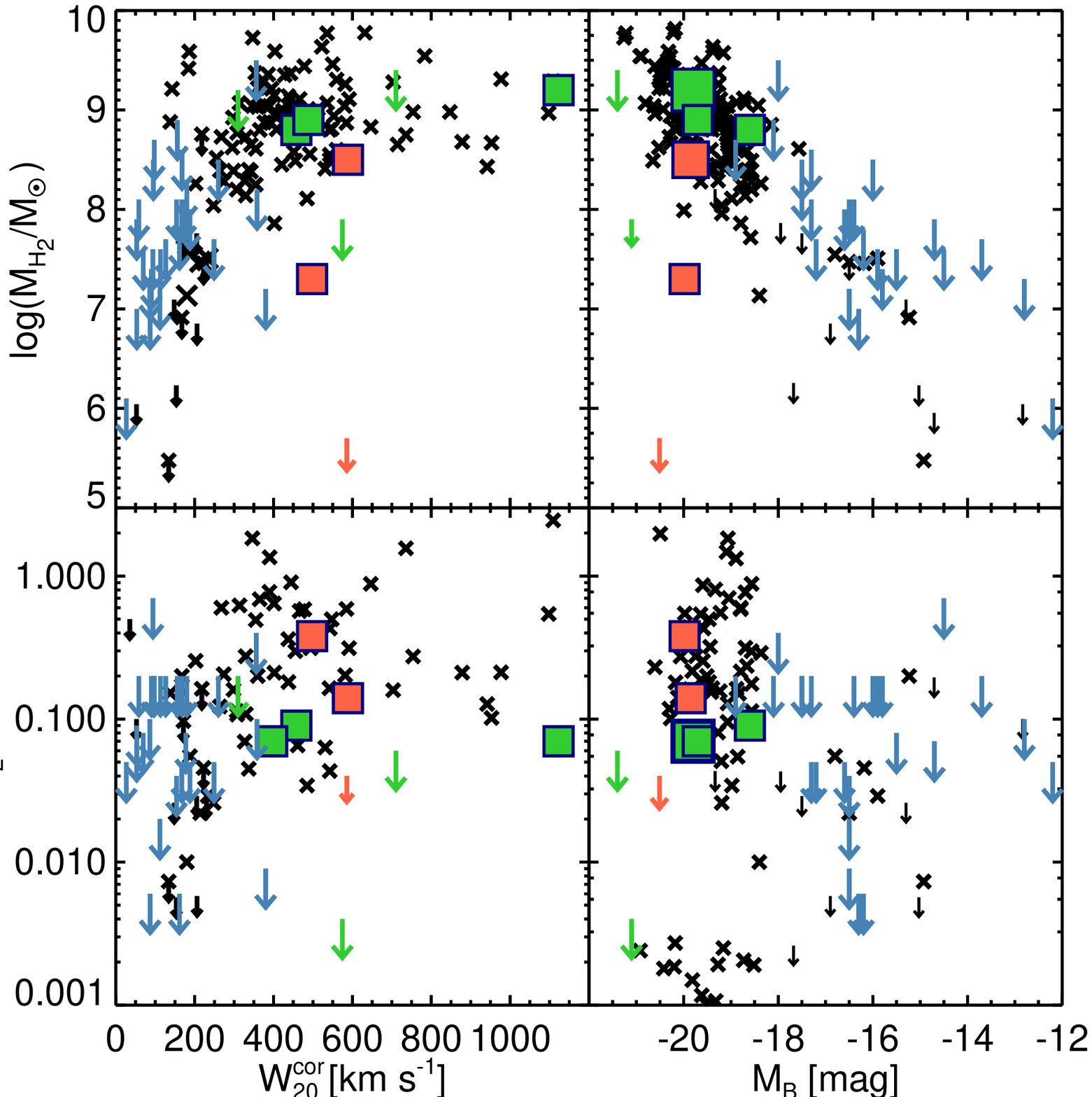}
\caption{Inclination corrected HI velocity widths versus
H$_2$ mass (top left) and the H$_2$-to-HI mass
ratio (bottom left).  At right is the absolute B magnitude versus H$_2$ mass
(top right) and the H$_2$-to-HI mass
ratio (bottom right).  The red, blue and green symbols are LSB galaxies 
and the black symbols are taken from various studies of the CO content in
HSB spiral galaxies. 
An arrow indicates only an upper limit was found.
See \citet{oneil04} for further information on this figure.\label{fig:CO} }
\end{figure}

Detection of molecular gas in LSB galaxies has been notoriously difficult.
In spite of attempts at detecting CO on LSB galaxies for more than 20 years 
\citep[e.g.][]{braine00, deblok98, schombert90} the first detection was only 
8 years ago \citep{oneil00}.  Since then a handful of CO detections have been made 
\citep[See][and references therein]{oneil04}, and in all cases the detections
have been in massive LSB systems.  However, comparing the LSB galaxy CO results with 
surveys of high surface brightness galaxies shows the  we find
the MLSB galaxies' M$_{H_2}$ and M$_{H_2}$/M$_{HI}$ values
fall within the ranges typically
found for high surface brightness objects, albeit at the low end of the distribution
(Figure~\ref{fig:CO}).

\section{Conclusions}
In summary, we clearly have a large number of known LSB galaxies which are fairly massive,
and this number is growing rapidly as more as more searches are undertaken.  As a result it
is finally becoming feasible to look at the galaxies as a class rather than just as individuals, 
and to try and apply what we learn to galaxy formation and evolution theories.

One interesting theory on the formation of massive LSB galaxies that was recently put forth is the
idea that massive LSB galaxies formed as the result of the collision of two galaxies \cite{mapelli08}.  This theory
can clearly explain a number of the massive LSB systems we have seen. such as UGC 06614 and possibly
Malin 1.  But it cannot explain all of the galaxies which we have found as the theory requires 
the galaxies {\it not} be undergoing any recent large star formation episodes, in clear contradiction
to many of the galaxies in out surveys.

\section{The $<$2 GHz Radio Future for Massive LSB Galaxies}
Over the next five years or so we clearly need to continue our HI surveys
 of the Universe to find and identify massive LSB galaxies.  To perform this 
searches we need both a large aperture telescope (for surface brightness sensitivity) and
also a high bandwidth to allow for searching a larger volume of the Universe at a given time.
The AGES survey (http://www.naic.edu/$\sim$ages) should provide just such a dataset, and we
are looking forward to seeing the final survey results.

Looking farther to the future the surveys which will be possible with, e.g. the Square Kilometer Array
will allow for an increase in the number of known massive LSB galaxies by factors of 100s or more
while simultaneously providing not only the total flux of the galaxy but information on the 
gas distribution of the galaxy and its nearby neighbors.  This level of information and sensitivity
should revolutionize the field.


\begin{theacknowledgments}
This work could not have been done without the help of Paul, Max, Willie
and a large number of collaborators, far too many
to thank individually. 

The National Radio Astronomy Observatory is a facility of the National Science Foundation operated under cooperative agreement by Associated Universities, Inc.  The Arecibo Observatory is part of the National Astronomy and Ionosphere Center, which is operated by Cornell University under a cooperative agreement with the National Science Foundation.  The Nan\c{c}ay Radio Observatory, which is the Unit\'e Scientifique de Nan\c{c}ay of the Observatoire de Paris, is associated with the French Centre National de Recherche Scientifique (CNRS) as USR B704, and acknowledges the financial support of the R\`egion Centre as well as of the European Union. This research also has made use of the Lyon-Meudon Extragalactic Database (LEDA), recently incorporated in HyperLeda, the NASA/IPAC Extragalactic Database (NED) which is operated by the Jet Propulsion Laboratory, California Institute of Technology, under contract with the National Aeronautics and Space Administration and the 
Sloan Digital Sky Survey which is managed by the Astrophysical Research Consortium for the Participating Institutions. The Participating Institutions are the American Museum of Natural History, Astrophysical Institute Potsdam, University of Basel, University of Cambridge, Case Western Reserve University, University of Chicago, Drexel University, Fermilab, the Institute for Advanced Study, the Japan Participation Group, Johns Hopkins University, the Joint Institute for Nuclear Astrophysics, the Kavli Institute for Particle Astrophysics and Cosmology, the Korean Scientist Group, the Chinese Academy of Sciences (LAMOST), Los Alamos National Laboratory, the Max-Planck-Institute for Astronomy (MPIA), the Max-Planck-Institute for Astrophysics (MPA), New Mexico State University, Ohio State University, University of Pittsburgh, University of Portsmouth, Princeton University, the United States Naval Observatory, and the University of Washington

\end{theacknowledgments}



\bibliographystyle{aipproc}   

\bibliography{mybib}

\IfFileExists{\jobname.bbl}{}
 {\typeout{}
  \typeout{******************************************}
  \typeout{** Please run "bibtex \jobname" to optain}
  \typeout{** the bibliography and then re-run LaTeX}
  \typeout{** twice to fix the references!}
  \typeout{******************************************}
  \typeout{}
 }

\end{document}